\documentclass[nojss]{jss} 

\author{Thomas Nagler \\Technische Universit\"at M\"unchen}
\title{\pkg{kdecopula}: An \proglang{R} Package for the Kernel Estimation of Bivariate Copula Densities}

\Plainauthor{Thomas Nagler} 
\Plaintitle{kdecopula: An R Package for the Kernel Estimation of Bivariate Copula Densities} 
\Shorttitle{\pkg{kdecopula}: An \proglang{R} Package for the Kernel Estimation of Bivariate Copula Densities} 

\Abstract{
  We describe the \proglang{R} package \pkg{kdecopula} (current version 0.9.0), which provides fast implementations of various kernel estimators for the copula density. Due to a variety of available plotting options it is particularly useful for the exploratory analysis of dependence structures. It can be further used for accurate nonparametric estimation of copula densities and resampling.
The implementation features spline interpolation of the estimates to allow for fast evaluation of density estimates and integrals thereof. We utilize this for a fast renormalization scheme that ensures that estimates are \emph{bona fide} copula densities and additionally improves the estimators' accuracy. The performance of the methods is illustrated by simulations.
}
\Plainkeywords{dependence, copula, nonparametric, kernel density, exploratory data analysis, R} 

\Address{
  Thomas Nagler\\
  Technische Universit\"at M\"unchen \\
Zentrum Mathematik \\
Lehrstuhl f\"ur Mathematische Statistik \\
Boltzmannstra\ss e 3, 
85748 Garching
}

\usepackage{setspace}

\usepackage{latexsym}
\usepackage{amssymb}
\usepackage{amsmath}
\usepackage{amsthm}
\usepackage{dsfont}
\usepackage{mathrsfs}
\usepackage{empheq}
\usepackage{bm}
\usepackage{subfig}
\usepackage{enumerate}
\usepackage{arydshln}

\newcommand{\ind}{\mathds{1}}
\newcommand{\R}{\mathds{R}}
\newcommand{\N}{\mathds{N}}

\newcommand{\wh}{\widehat}

\begin{document}

\section{Introduction}

Dependence modeling with copulas has attracted a lot of attention in recent decades. By now, copulas are established tools in many fields of applied statistics, such as finance \citep{Cherubini04}, hydrology \citep{Salvadori07}, or machine learning \citep{Elidan13}.

At the very heart of copula theory is the famous theorem of Sklar \citep{Sklar59}. It states that any multivariate distribution function can be decomposed into the marginal distributions and a copula, which captures the dependence between variables. Let $X$ and $Y$ be two continuous random variables with joint distribution $F$ and marginal distributions $F_X$ and $F_Y$. Then, for all $(x, y)$ in the support of the random vector $(X,Y)$,
\begin{align*}
F(x,y) = C\bigl(F_X(x), F_Y(y)\bigr).
\end{align*}
The copula $C\colon [0,1]^2 \to [0,1]$ is the bivariate distribution function of the random vector $(U, V) =  \bigl(F_X(X), F_Y(Y)\bigr)$ which has uniform marginal distribution. If $F$ admits a density, we can also decompose the density $f$ into
\begin{align}
	f(x,y) = c\bigl(F_X(x), F_Y(y)\bigr)f_X(x)f_Y(y), \label{introduction:sklar_eq}
\end{align} 
where $c$, $f_X$ and $f_Y$ are the densities corresponding to $C$, $F_X$ and $F_Y$, respectively.

One of the major benefits of copula-based modeling is that inference for marginal distributions can be separated from the modeling of the dependence structure, i.e., the copula. For the estimation of the copula density $c$, it is most common to take a two-step approach: First, obtain estimates $\wh F_X, \wh F_Y$ of the marginal distributions. A convenient and flexible way to do this is to use the empirical distribution function as an estimator. Second, define \emph{pseudo-observations} $\bigl(\wh U, \wh V\bigr) =  \bigl(\wh F_X(X), \wh F_Y(Y)\bigr)$. The copula density is then estimated as the joint density of $\bigl(\wh U, \wh V\bigr)$. 

Often, one assumes a parametric model for the copula density $c$ and estimates its parameters by maximum-likelihood. Although there is a large variety of parametric copula models, they notoriously lack flexibility and bear the risk of misspecification. Nonparametric density estimators remedy these issues. But since copulas live on a bounded support --- the unit hypercube --- estimators have to be carefully tailored to this problem. 

A specific class of nonparametric density estimators are kernel estimators. They are a popular tool for exploratory data analysis and widely used in many disciplines \citep[e.g.,][]{Aitken04, Kie10}. The package \pkg{kdecopula} implements several bivariate kernel copula density estimators that have been proposed in recent years. In a nutshell, the package provides methods for:
\begin{itemize}
	\item estimation,
    \item bandwidth selection,
    \item simulation,
    \item visualization.
\end{itemize}

There exist two alternative methods for the kernel estimation of copula densities in \proglang{R} \citep{R}: the function \code{kcopula} from the \pkg{ks} package \citep{ks} implements an estimator that glues together two independent kernel estimates for the center and boundary of the unit square; the function \code{npcopula} from the \pkg{np} package \citep{np} derives the copula density from an estimate of the joint distribution \citep[as proposed by][]{racine2015mixed}. However, these implementation do not reflect the numerous specialized contributions on the topic. Our package closes this gap by implementing state-of-the-art methods for kernel copula density estimation. The implemented methods are substantially more accurate than existing implementations. Additionally, the package provides a normalization algorithm which ensures that estimates are a \emph{bona fide} copula densities and further improves the accuracy. 

Apart from kernel estimators, \citet{pencopula} implemented nonparametric copula density estimators based on penalized likelihood estimation in the \pkg{pencopula} package (using B-splines or Bernstein polynomials). The extension \pkg{penDvine} \citep{penDvine} provides a convenient version with automatic bandwidth selection. A comparison with our implementation will show that these estimators are only competitive when the dependence is weak. The author is not aware of any software implementations for nonparametric copula density estimation outside of \proglang{R}.

In \autoref{sec:review}, we give a review of kernel copula density estimators and point to the relevant literature. \autoref{sec:functionality} describes the functionality of the package and gives examples for its use. In \autoref{sec:splines}, we give background on the implementation of the estimators using spline interpolation for fast evaluation and renormalization of the estimates. The statistical accuracy of the estimators in this package and other nonparametric copula density estimators (see previous paragraphs) is compared in \autoref{sec:simulations}.  A summary is given in \autoref{sec:conclusion}.
\section{Kernel estimators of the copula density: a review}
\label{sec:review}
This section will review different approaches to kernel estimation of the copula density. As is common in the literature, we focus on the bivariate case.

Assume we have \emph{iid} observations $(U_i, V_i)$, $i = 1, \dots, n$, from a bivariate copula $C$ and are interested in the estimation of the corresponding density $c(u,v)$. One could apply the usual kernel density estimator to this problem:
\begin{align*}
\wh c_n(u,v) = \frac 1 {n} \sum_{i=1}^{n} K_{b_n}\bigl(u-U_i\bigr) K_{b_n}\bigl(v-V_i  \bigr), \qquad  (u,v) \in [0,1]^2,
\end{align*}
where we used the notation $K_{b} (\cdot) = K (\cdot /{b})/b$. The kernel function $K$ is typically assumed to be a symmetric, bounded probability density function on $\R^2$ and $b_n>0$ is the smoothing or bandwidth parameter. 
There is a problem, however. The estimator will put a considerable amount of probability mass outside of the unit square. This implies that $\wh c_n$ is not a density function on $[0,1]^2$, because it does not integrate to one. The estimator will additionally suffer from severe bias at the boundaries \citep[see, e.g.,][]{Charpentier06}. Three different approaches to tackle this problem have emerged. All three techniques arose initially in the context of univariate kernel density estimation on the unit line. The following sections explain the ideas behind them (in the context of copulas) and give references for more detailed accounts.

\subsection{The mirror-reflection method}

\begin{figure}[t]
\centering
\subfloat[Original data]{\includegraphics[width=0.4\textwidth]{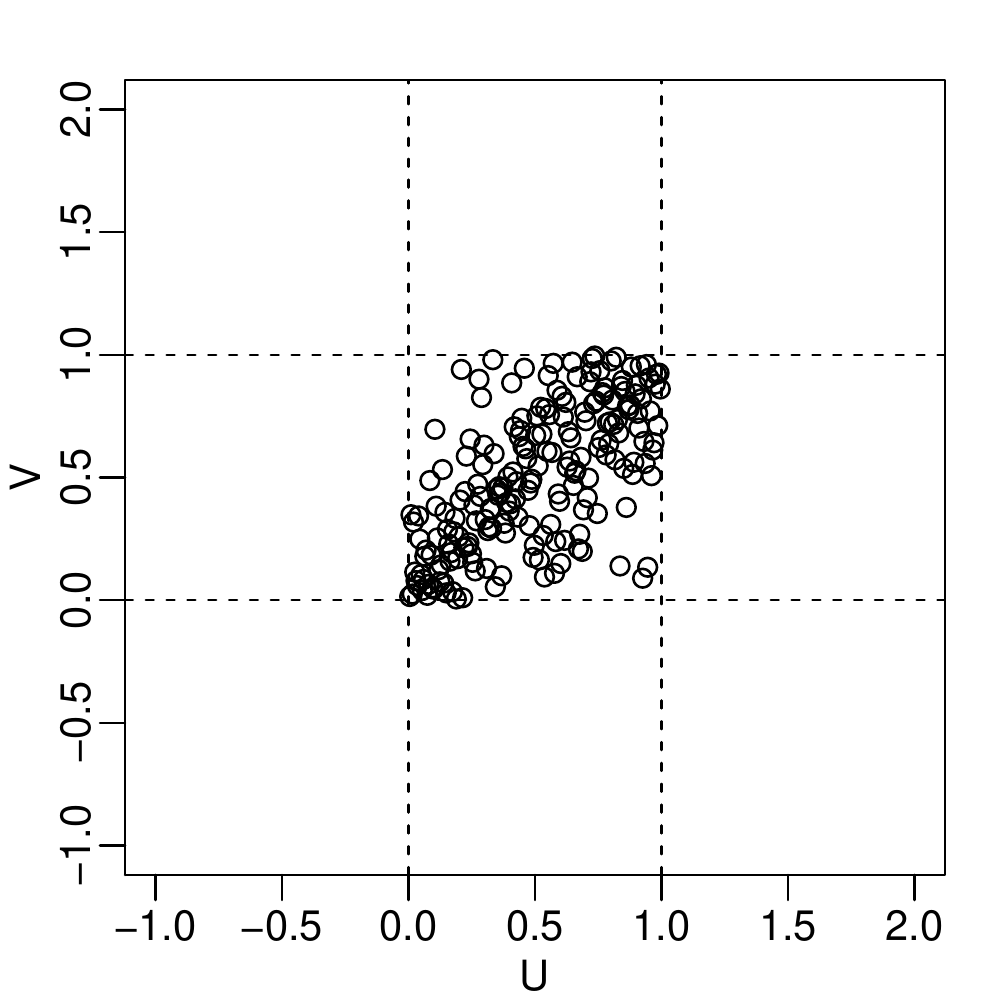}} \hspace{0.1\textwidth}
\subfloat[Augmented data]{\includegraphics[width=0.4\textwidth]{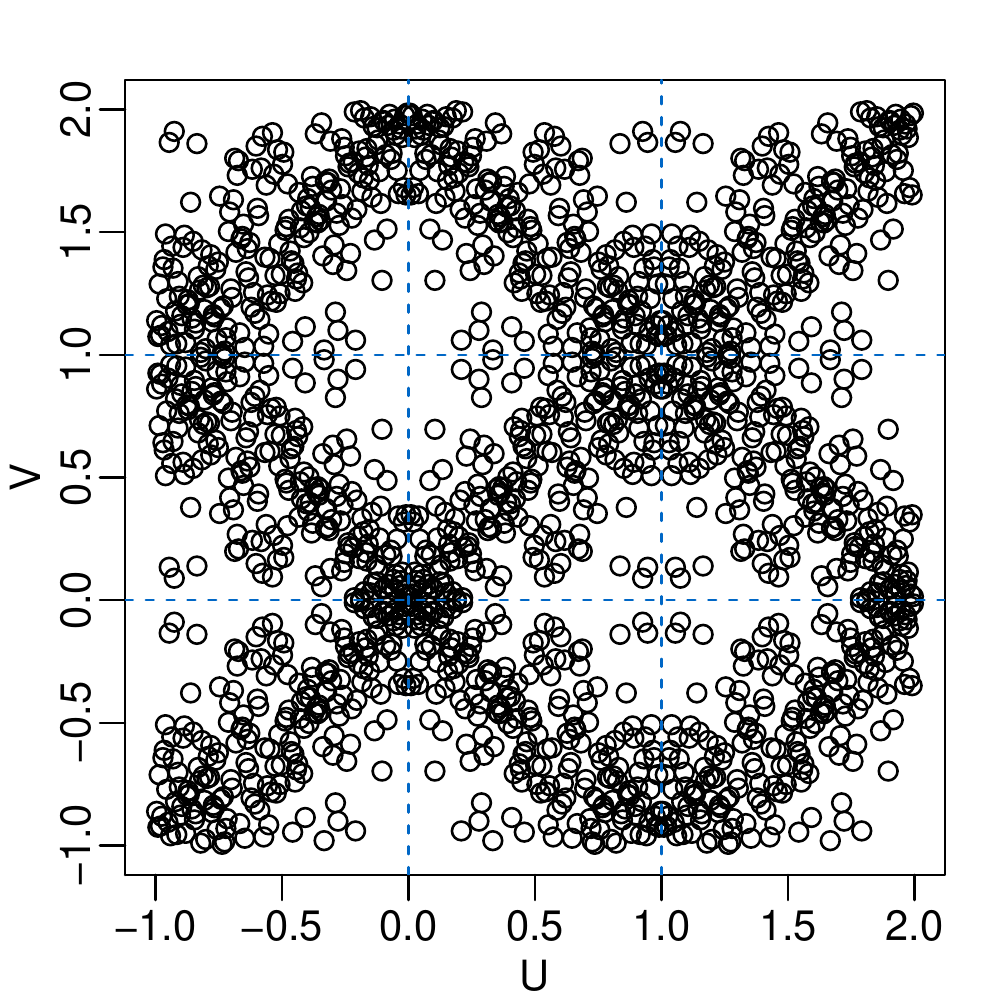}}
\caption{The data augmentation process. The set in (b) is obtained by reflecting all original data points w.r.t.\ to all corners and edges.}
\label{fig:mr_data}
\end{figure} 

An intuitive way of adapting $\wh c_n$ to make sure that it is a density on $[0,1]^2$ is the following: gather all probability mass that was put outside of the unit square, and redistribute it back to $[0,1]^2$. This is the idea behind the \emph{mirror-reflection technique}, which was proposed for copula density estimation by  \citet{Gijbels90}. As indicated by the name, all data are reflected at the corners and edges of the boundary region. The augmented data set containing all reflections is given by
\begin{align*}
\bigl(\tilde U_{ik}, \tilde V_{ik}\bigr)_{k=1, \dots,9} &= \biggl\{ (U_i, V_i), (-U_i,V_i), (U_i, -V_i), (-U_i, -V_i), (U_i,2-V_i), \\
& \phantom{= \biggl\{}(-U_i, 2-V_i), (2-U_i, V_i), (2 - U_i, -V_i), (2- U_i, 2-V_i) \biggr\}.
\end{align*}
A visualization of the augmented data set is given in \autoref{fig:mr_data}. The \emph{mirror-reflection estimator} is then defined as the usual kernel density estimator on the augmented data:
\begin{align*}
\wh c_n^{(MR)}(u,v) = \frac 1 {n} \sum_{i=1}^{n} \sum_{k=1}^{9} K_{b_n}\bigl(u-\tilde U_{ik}\bigr) K_{b_n}\bigl(v- \tilde V_{ik}  \bigr), \qquad  (u,v) \in [0,1]^2.
\end{align*}
By reflecting all data points at the corners and edges also the probability mass outside of the unit square gets reflected back to the interior. As a result, the estimator now integrates to one.
A detailed analysis of the asymptotic properties and a method for automatic bandwidth selection are given in \citet{Nagler14}.

\subsection{The beta kernel method}

A second approach is to use kernels whose support matches the  support of the density we want to estimate, and vary the shape of those kernels depending on the point where density shall be estimated. This is achieved by so-called \emph{boundary kernels}, and \emph{beta kernels} are one instance. An estimator of the copula density based on this idea was proposed by \citet{Charpentier06}:
\begin{align*}
c^{(\beta)}(u,v) = \frac 1 {n} \sum_{i=1}^{n} \beta\biggl(U_i; \frac u {b_n} +1,\frac {1-u} {b_n} +1 \biggr)\beta\biggl(V_i; \frac v {b_n} +1,\frac {1-v} {b_n} +1 \biggr), \qquad (u,v) \in [0,1]^2,
\end{align*}
where $\beta(\cdot;p,q)$ is the density of a $\mbox{Beta}(p,q)$-distributed random variable. We refer to \citet{Nagler14} for details on asymptotics and bandwidth selection.

\subsection{The transformation method}

A third approach is inspired by the early work of \citet{Devroye85} and was introduced to kernel copula density estimation by \cite{Charpentier06}. The simple idea is to transform the data so that it is supported on the full $\R^2$ (instead of the unit cube). On this transformed domain, standard kernel techniques can be used to estimate the density. An adequate back-transformation then yields an estimate of the copula density. For the transformation, the inverse of standard normal $cdf$ is most common since it is known that kernel estimators tend to do well for Gaussian random variables.

\begin{figure}[t]
\centering
\subfloat[]{\includegraphics[width=0.225\textwidth]{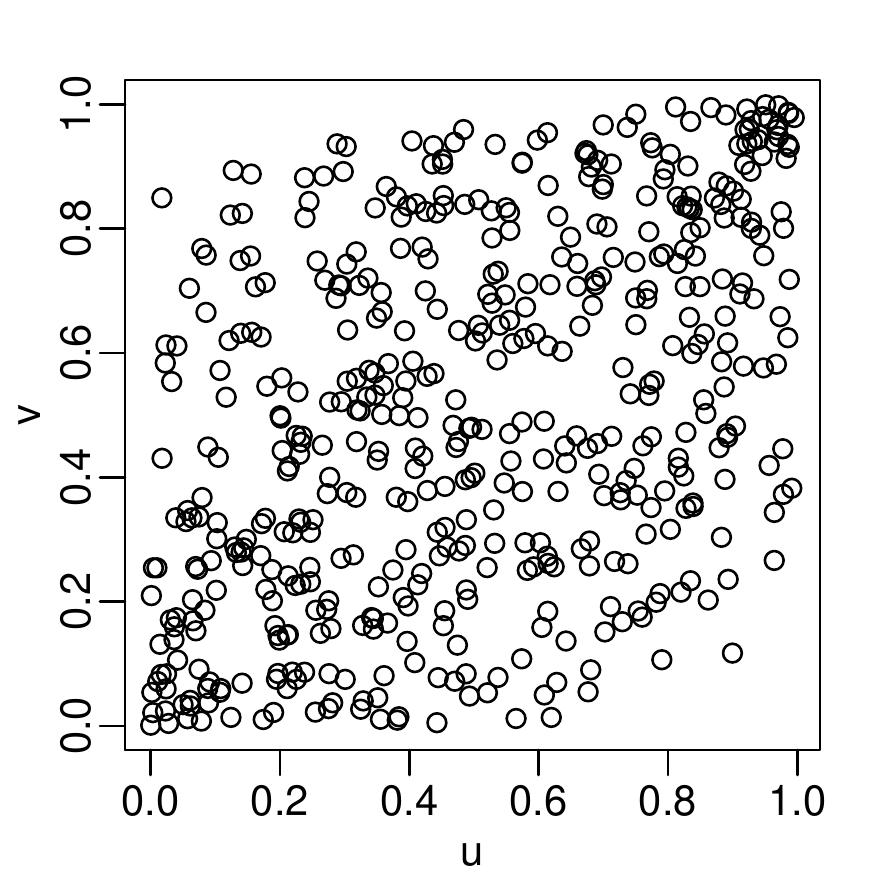}}
\subfloat[]{\includegraphics[width=0.225\textwidth]{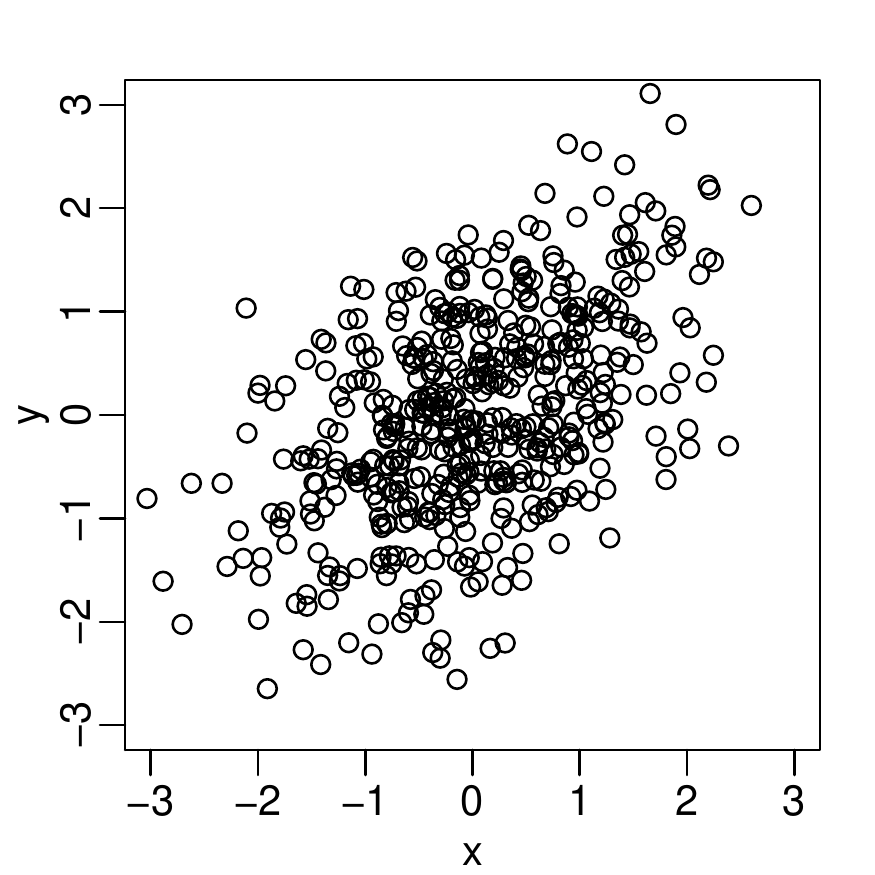}}
\subfloat[]{\includegraphics[width=0.25\textwidth]{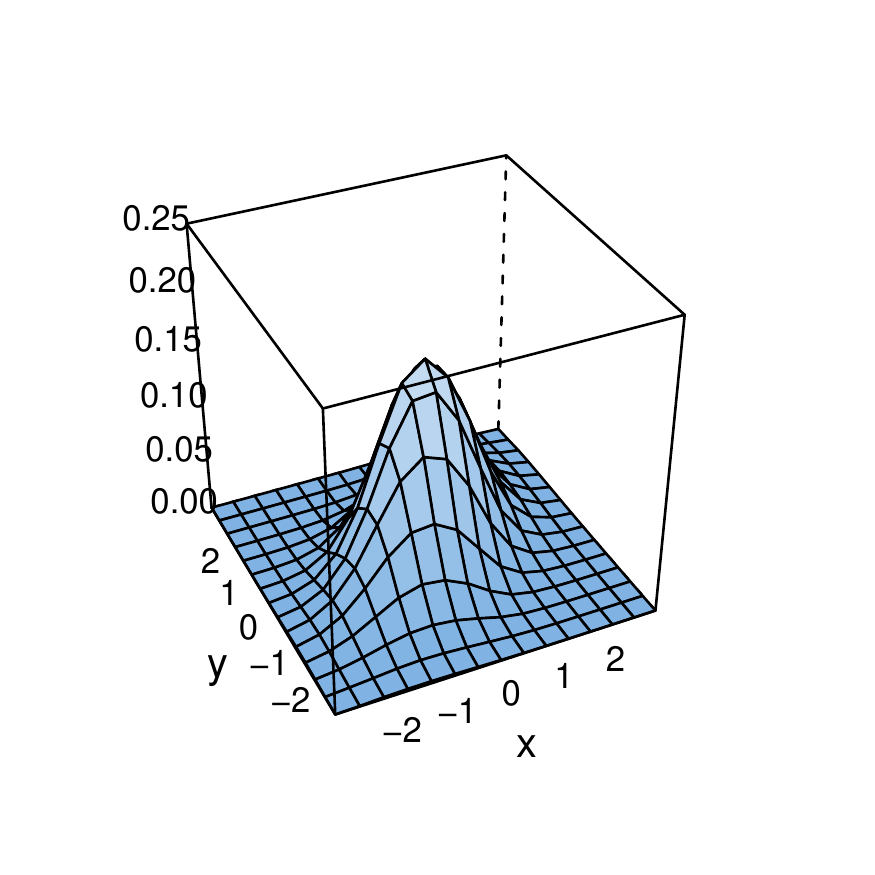}}
\subfloat[]{\includegraphics[width=0.25\textwidth]{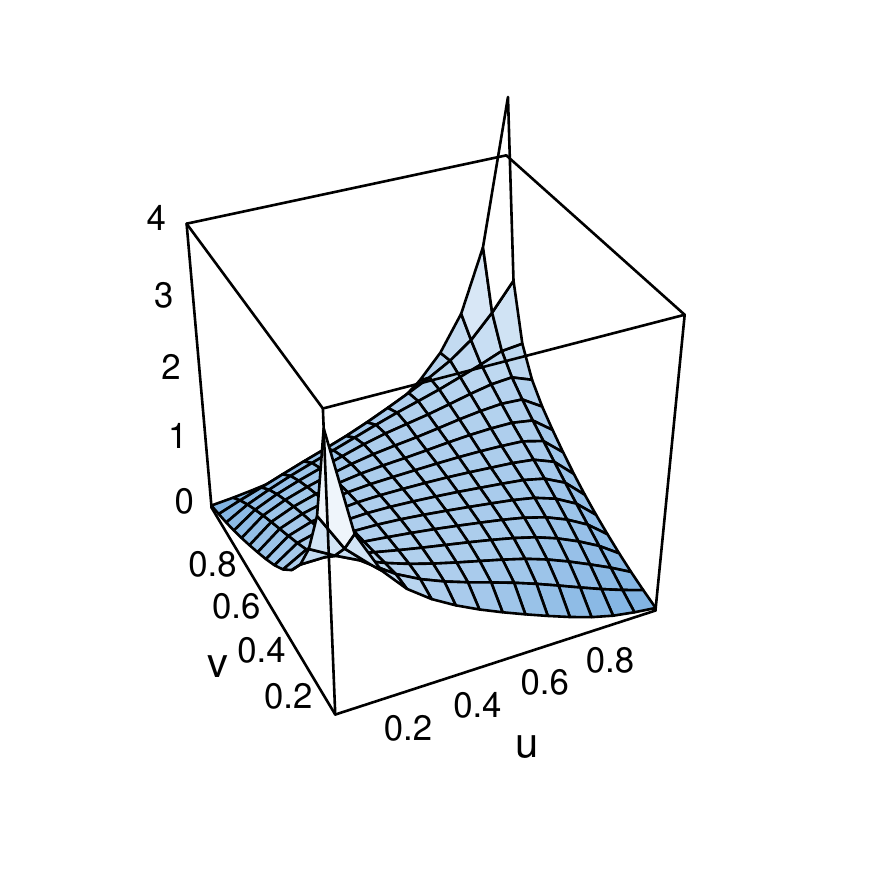}}
\caption{The transformation estimator. (a) copula sample; (b) transformed sample;  (c) kernel density estimate for transformed sample; (d) kernel estimate of the copula density.}
\label{fig:trafo}
\end{figure}

Denote $\Phi$ as the standard Gaussian $cdf$ and $\phi$ its first order derivative. Then $(X_i,Y_i) = \bigl(\Phi^{-1}(U_i), \Phi^{-1}(V_i)\bigr)$ is a random vector with Gaussian margins and copula $C$. By Sklar's Theorem, the corresponding density $f$ can be written as
\begin{align}
f(x,y) = c\bigl(\Phi(x), \Phi(x)\bigr) \phi(x) \phi(y).
\label{eq:trafo_sklar}
\end{align}
This density can be easily estimated by a standard kernel estimator. From such an estimator $\wh f_n$, we can derive an estimator for the copula density $c$ by isolating $c$ in \eqref{eq:trafo_sklar}:
\begin{align}
\wh c_n^{(T)}(u,v) = \frac { \wh f_n\bigl(\Phi^{-1}(u), \Phi^{-1}(v)\bigr) }{\phi\bigl(\Phi^{-1}(u)\bigr)\phi\bigl(\Phi^{-1}(v)\bigr)}, \qquad  (u,v) \in [0,1]^2. \label{eq:trafo_def}
\end{align}
This procedure is illustrated in \autoref{fig:trafo}. The left panel shows the original data from the copula density $c$; next to it we see the transformed data after the inverse Gaussian $cdf$ has been applied. The third plot shows a kernel estimate of the density of the transformed data; and finally, the fourth plot shows the corresponding kernel estimate of the copula density.

The most natural choice for $f_n$ is the conventional kernel density estimator. More recently, \citet{Geenens14a} proposed to use a local likelihood estimator with nearest-neighbor bandwidths instead. Another recent extension was introduced by \citet{Wen15} who suggested to taper the back-transformation in the tails by increasing the variance of the Gaussian densities in the denominator of \eqref{eq:trafo_def}. For more details, we refer to the original papers.

\section{The package's functionality}
\label{sec:functionality}

In the following, we describe the most important functions provided by the package. All function either produce or take objects of the S3-class \code{kdecopula} for which several methods are available.


\subsection[Estimation and bandwidth selection: kdecop]{Estimation and bandwidth selection: \code{kdecop}}

At the core of the \pkg{kdecopula} package is the function \code{kdecop}, which estimates the copula density from data. The only mandatory input is an $n \times 2$ matrix of copula data, i.e., data with standard uniform margins. Such data is usually obtained in a first step by applying the empirical marginal $cdf$s to the data. This is equivalent to a rank transformation as shown below. The following lines of code load the package and an accompanying data set. The data is transformed to uniform margins in the third line, and the last line fits the kernel estimator.
\begin{CodeChunk}
\begin{CodeInput}
R> library("kdecopula")
R> data("wdbc")
R> uv <- apply(wdbc[, c(2, 8)], 2, rank) / (nrow(wdbc) + 1)
R> fit <- kdecop(uv)
\end{CodeInput}
\end{CodeChunk}
The output of the function \code{kdecop} is an object of class \code{kdecopula} that contains all information collected during the estimation process. The most relevant information can be summarized as follows.
\begin{CodeChunk}
\begin{CodeInput}
R> summary(fit)
\end{CodeInput}
\begin{CodeOutput}
Kernel copula density estimate (tau = 0.47)
------------------------------
Variables:    mean radius -- mean concavity 
Observations: 569 
Method:       Transformation local likelihood, log-quadratic (nearest-neighbor,
  'TLL2nn') 
Bandwidth:    alpha = 0.3519647
              B = matrix(c(0.71, 0.7, -0.7, 0.71), 2, 2)
---
logLik: 201.22    AIC: -367.97    cAIC: -366.83    BIC: -293.11 
Effective number of parameters: 17.23
\end{CodeOutput}
\end{CodeChunk}
Summary statistics such as \emph{AIC} or the \emph{effective number of parameters/degrees of freedom} can be accessed via the usual generic functions.
\begin{CodeChunk}
\begin{CodeInput}
R> logLik(fit)
\end{CodeInput}
\begin{CodeOutput}
'log Lik.' 201.2196 (df=17.23373)
\end{CodeOutput}
\begin{CodeInput}
R> AIC(fit)
\end{CodeInput}
\begin{CodeInput}
[1] -367.9718
\end{CodeInput}
\end{CodeChunk}

The function \code{kdecop} provides all estimation methods mentioned in \autoref{sec:review}. The estimation method can be specified via the \code{method} argument, e.g., \code{kdecop(..., method = "MR")}. For each method, we have implemented an automatic bandwidth selection procedure. Below we list all implemented methods including a reference to the bandwidth selection procedure used:
\begin{description}
\item[\code{MR}] \hfill \\
The mirror-reflection estimator of \citet{Gijbels90}. Smoothing parameters are selected by minimizing the AMISE using the Frank copula as the reference copula \citep[see][Section 3.2.4]{Nagler14}.

\item[\code{beta}] \hfill \\
The beta kernel estimator of \citet{Charpentier06}. Smoothing parameters are selected by minimizing the AMISE using the Frank copula as the reference copula  \citep[see,][Section 3.3.3]{Nagler14}.

\item[\code{T}]  \hfill \\
The transformation estimator of \citet{Charpentier06}, but allowing for a bandwidth matrix and not just one parameter. The bandwidth matrix is set by a rule of thumb which is the normal reference rule on the transformed domain \citep[see,][Section 3.4.4]{Nagler14}:
\begin{align*}
B_{\mathrm{rot}} = n^{-1/6} \widehat{\Sigma}_{\bm Z}^{1/2},
\end{align*}
where $\widehat{\Sigma}_{\bm Z}$ is the empirical covariance matrix of $\Phi^{-1}(U_i)$ and $\Phi^{-1}(V_i)\bigr)$, $i = 1, \dots, n$. 

\item[\code{TLL1}, \code{TLL2}, \code{TLL1nn}, \code{TLL2nn} (default)]  \hfill \\
The transformation local likelihood estimator of \citet{Geenens14a}. \code{TLL1} approximates the log-density linearly; \code{TLL2} by quadratic polynomials. The -\code{nn} versions use nearest-neighbor bandwidths instead of fixed ones. For fixed-bandwidth versions, the bandwidth matrix is set by the rule of thumb
\begin{align*}
B_{\mathrm{rot}} = 3 n^{-1/(4q^* + 2)} \widehat{\Sigma}_{\bm Z}^{1/2}, \quad q^* = 1 + \lfloor q/2 \rfloor,
\end{align*}
where $q$ is the degree of the polynomial, $\widehat{\Sigma}_{\bm Z}$ is the empirical covariance matrix of $\Phi^{-1}(U_i)$ and $\Phi^{-1}(V_i)\bigr)$, $i = 1, \dots, n$. This rule of thumb is similar to the normal reference rule, but ensures that the bandwidth matrix vanishes at the mean-square optimal rate.
For nearest-neighbor methods, smoothing parameters are selected based on univariate least-squares cross-validation on the first principal component in the transformed domain \citep[see,][Section 4]{Geenens14a}. Local likelihood fitting is implemented via the \pkg{locfit} package \citep{locfit}.

\item[\code{TTCV}, \code{TTPI}]  \hfill \\
Tapered transformation estimator of \citet{Wen15}\footnote{The implementation of the tapered transformation estimators was kindly provided by Kuangyu Wen.}. Smoothing parameters are selected in the transformed domain by profile cross-validation \citep[\code{TTCV}, see,][Section 4.2]{Wen15} or plug-in minimization of the AMISE \citep[\code{TTPI}, see,][Section 4.1]{Wen15}.
\end{description}
It is possible to specify the bandwidths manually using the \code{bw} argument of \code{kdecop}, although we recommend against it. If it is necessary to manually make an estimate more or less smooth, we advise to use the bandwidth multiplier argument \code{kdecop(..., mult = 1)}. Values larger than one will make the estimate smoother; values less than one make the estimate less smooth.


\subsection[Working with the estimated density: (d/p/r)kdecop]{Working with the estimated density: \code{(d/p/r)kdecop}}

In analogy to the usual \code{(d/p/r)}-prefixes for distribution families in \proglang{R}, we provide \code{(d/p/r)}-versions for the \code{kdecop}-family. The functions \code{dkdecop} and \code{pkdecop} can be used to evaluate the density and \emph{cdf} of a \code{kdecopula} object, respectively.

\begin{CodeChunk}
\begin{CodeInput}
R> dkdecop(c(0.1, 0.2), fit) 
\end{CodeInput}
\begin{CodeOutput}
[1] 1.691764
\end{CodeOutput}
\begin{CodeInput}
R> pkdecop(cbind(c(0.1, 0.9), c(0.1, 0.9)), fit)
\end{CodeInput}
\begin{CodeOutput}
[1] 0.0327257 0.8505370
\end{CodeOutput}
\end{CodeChunk}

The \code{rkdecop} function simulates data from the estimated density. This can be done in two ways: a) using pseudo-random numbers based on \code{runif}, b) using quasi-random numbers based on \code{ghalton} from the \pkg{qrng} package \citep{qrng}.

\begin{CodeChunk}
\begin{CodeInput}
R> pseudo <- rkdecop(500, fit)
R> quasi  <- rkdecop(500, fit, quasi = TRUE)
\end{CodeInput}
\end{CodeChunk}


\subsection[Visualization: the plot and contour generics]{Visualization: the \code{plot} and \code{contour} generics}

\begin{figure}
	\subfloat[surface plot]{\includegraphics[width = 0.32\textwidth]{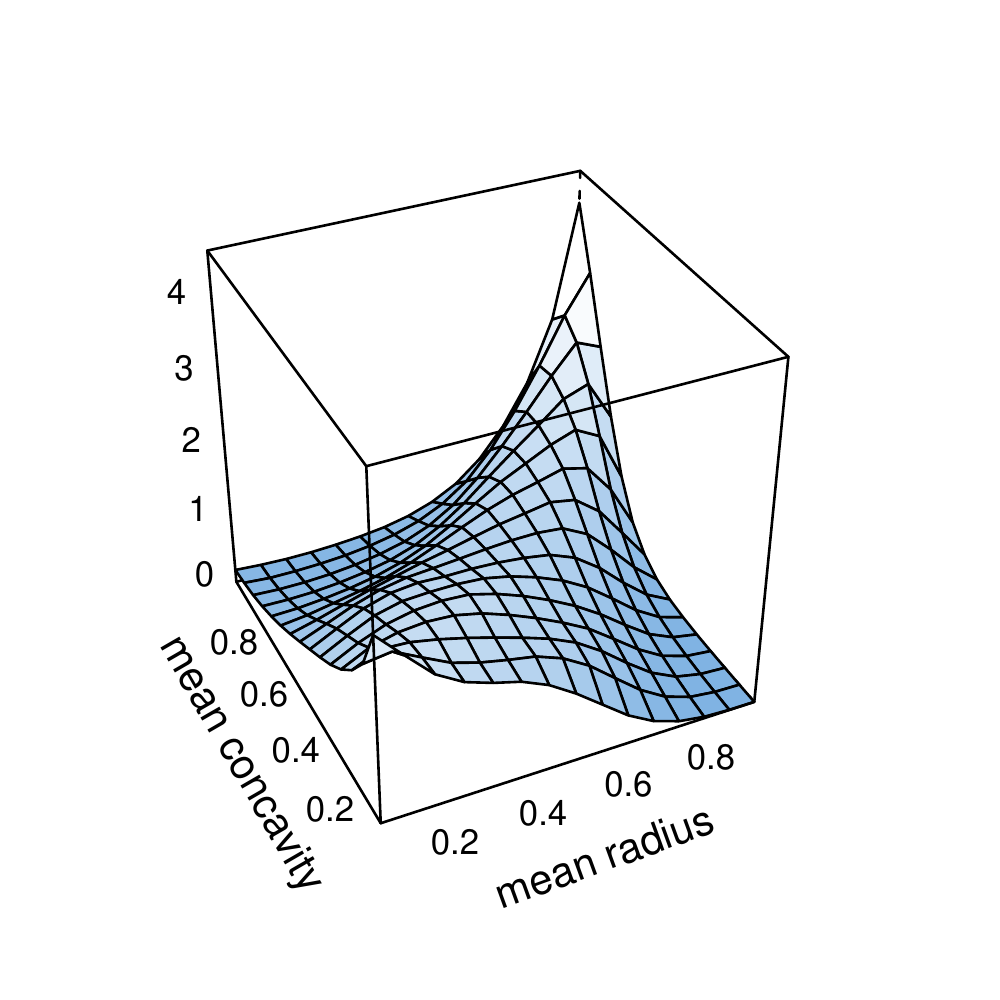}}
    \subfloat[contour plot]{\includegraphics[width = 0.32\textwidth]{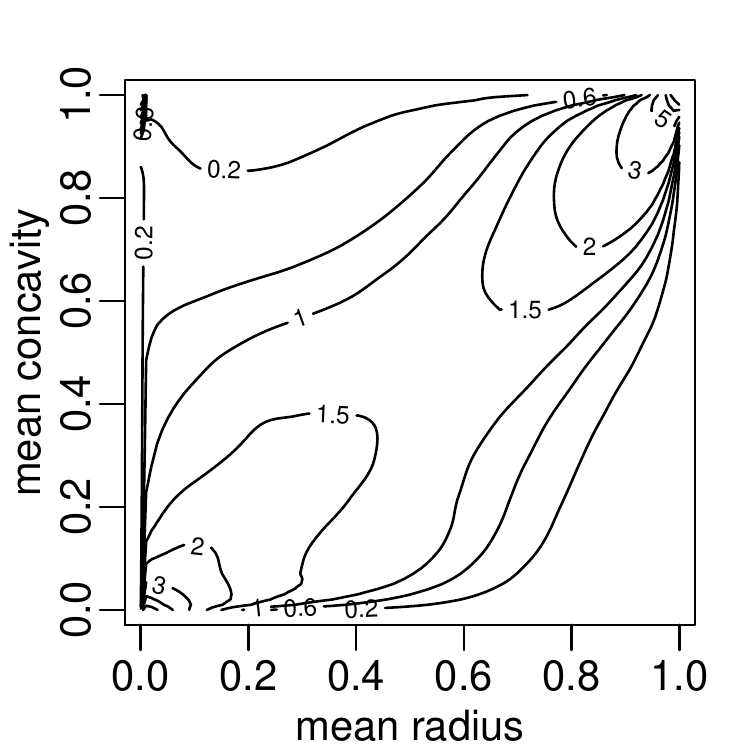}}
	\subfloat[marginal normal contour plot]{\includegraphics[width = 0.32\textwidth]{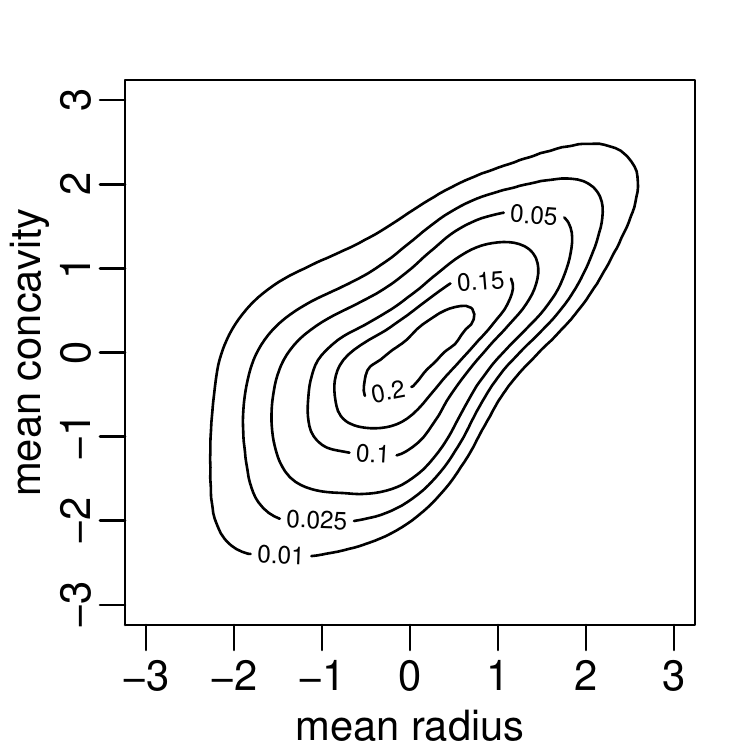}}
   \caption{Three ways to visualize a copula density.}
   \label{fig:plot_types}
\end{figure}

For many people, the most interesting feature is probably to make exploratory plots. There are three common ways to visualize a copula density: (a) a surface (or perspective) plot of the copula density, (b) a contour plot of the copula density, (c) a contour plot of the copula density when combined with standard normal margins. The following three lines of code produce the plots shown in \autoref{fig:plot_types}. Optionally, further arguments can be passed to improve the aesthetics.
\begin{CodeChunk}
\begin{CodeInput}
R> plot(fit)
R> contour(fit, margins = "unif") 
R> contour(fit) 
\end{CodeInput}
\end{CodeChunk}
\begin{figure}[t]
	\centering
    \subfloat[Independence]{\includegraphics[width = 0.3\textwidth]{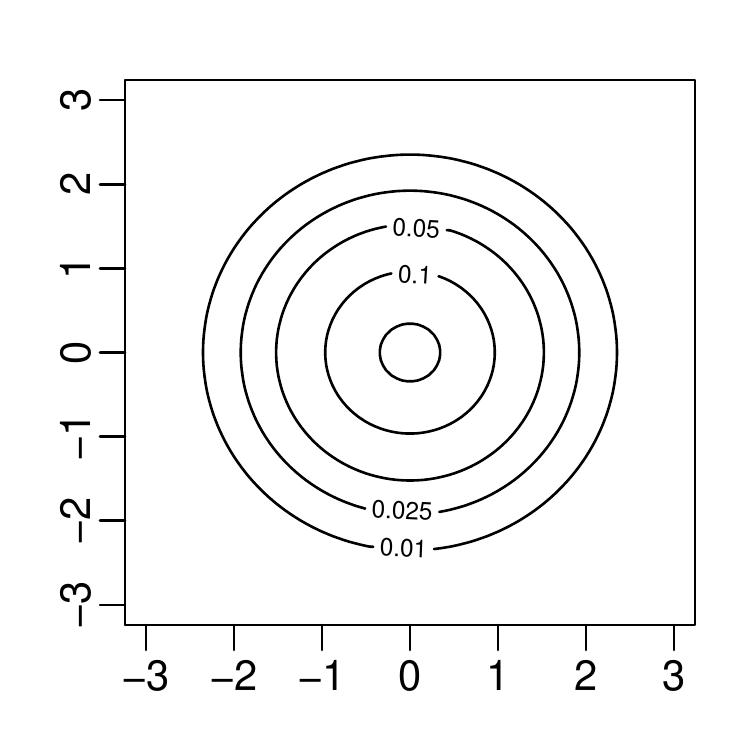} \label{fig:contours_indep}} \hspace{0.1cm}
    \subfloat[Gaussian]{\includegraphics[width = 0.3\textwidth]{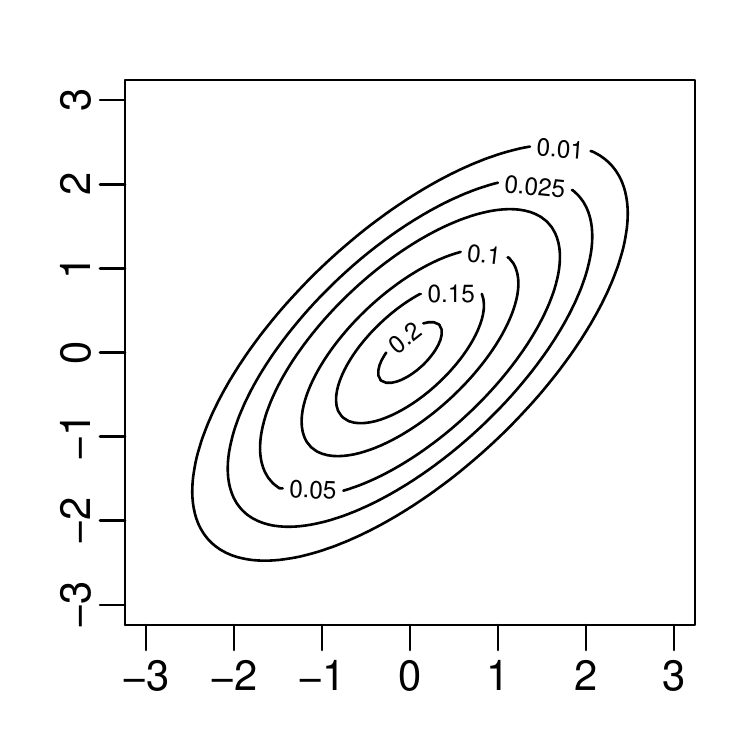} \label{fig:contours_gauss}} \hspace{0.1cm}
    \subfloat[Student t]{\includegraphics[width = 0.3\textwidth]{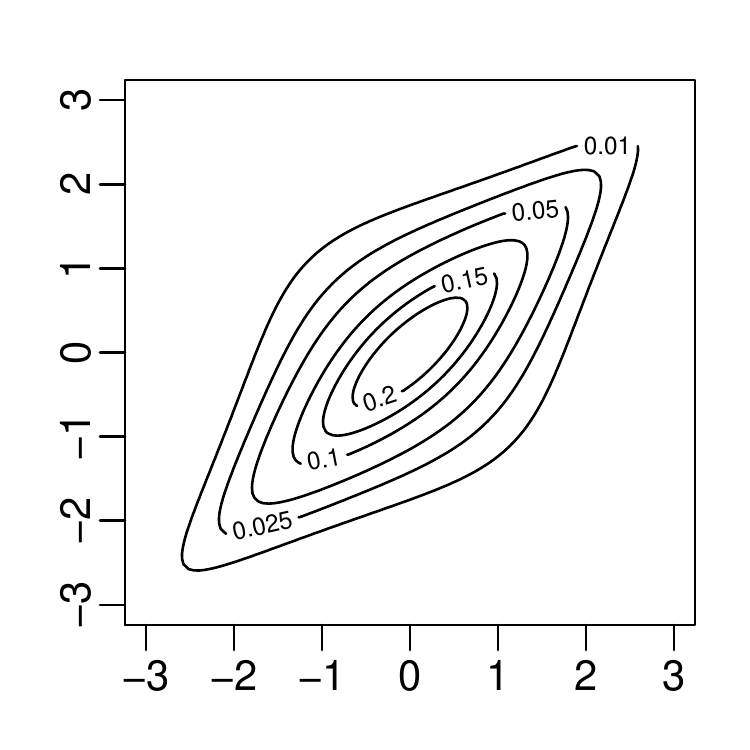} \label{fig:contours_t}} \\
    \subfloat[Gumbel]{\includegraphics[width = 0.3\textwidth]{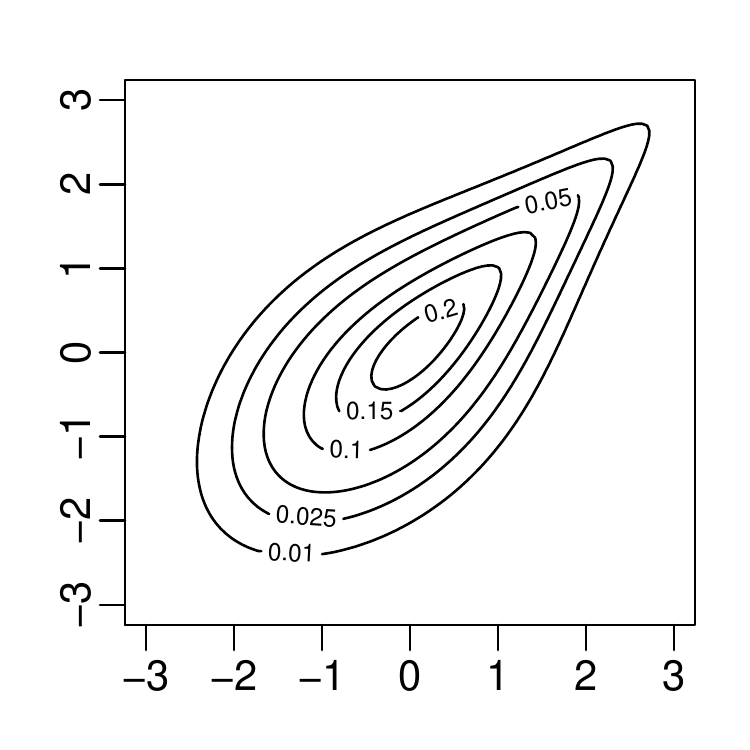} \label{fig:contours_gumbel}} \hspace{0.1cm}
    \subfloat[Clayton]{\includegraphics[width = 0.3\textwidth]{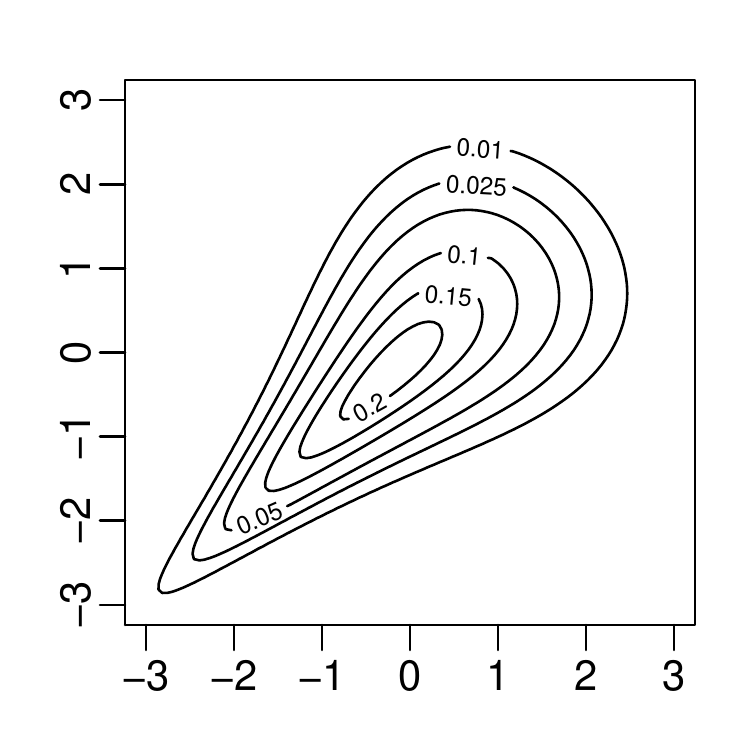} \label{fig:contours_clay}} \hspace{0.1cm}
    \subfloat[Frank]{\includegraphics[width = 0.3\textwidth]{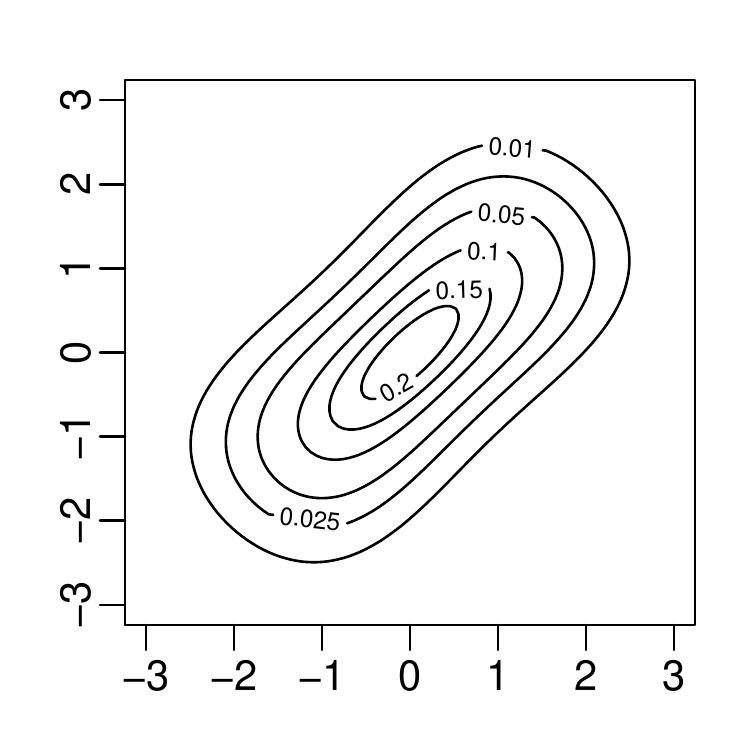} \label{fig:contours_frank}} \\
    \caption{Marginal normal contour plots for several parametric copula families.}
    \label{fig:contours}
\end{figure}

In the author's experience, the most useful plot is (c), the marginal normal contour plot. Copula densities usually explode at some corners of the unit square. As a result, it is not possible to reasonably visualize the raw density (a) on the whole $[0, 1]^2$. The function \code{plot.kdecopula()} therefore restricts the displayed area to $[0.01, 0.99]^2$. Obviously, this hides some information in the tails. This can be problematic because the tails are often of particular interest in copula models. But even on the restricted domain the copula density often attains values larger than 20 when there is strong dependence in the tails. On this scale, the shape of the copula density close to the center of the unit square is difficult to assess. Similarly, the contours of the raw density (b) are inappropriate to reflect the tail behavior because contour lines near to the corners become too close to be visually distinguishable. The marginal normal contour plot overcomes these issues by transforming the margins such that the transformed density is bounded. It has the additional advantage that it allows for an intuitive interpretation that is relative to the Gaussian copula as explained in the following paragraph.

If the true copula is the independence copula, the contours are perfect circles (see, \autoref{fig:contours_indep}). This is obviously not the case for the estimated density in \autoref{fig:plot_types}. \autoref{fig:contours_gauss} shows a Gaussian copula with Kendall's $\tau$ set to the estimated $\tau$ from the data. A Gaussian copula combined with Gaussian margins results in a bivariate Gaussian density and its contours are ellipses. Since most statisticians are familiar with this kind of distribution it seems natural to use this as a benchmark when interpreting the marginal normal contour plot for other copulas. The next plot, \autoref{fig:contours_t}, shows the Student t copula ($df = 3$). Here the contours look like a diamond due to the higher density values in the tails (i.e., the corners of the square). This reflects that --- in contrast to the Gaussian copula --- the Student t copula exhibits \emph{tail dependence}  \citep[e.g.,][]{Joe14}, a concept that is very important in the modeling of risks. In general, a spiky shape in the corners of the contours is an indication of tail dependence in the respective corner. This can be observed again in \autoref{fig:contours_gumbel} and \autoref{fig:contours_clay}, where the Gumbel and Clayton copula are shown, respectively. The Gumbel copula is asymmetric and features  upper tail dependence only. This is reflected by a spiky shape in the upper right corner and a flatter shape in the lower left corner. For the Clayton copula it is the other way around. Finally the Frank copula has no tail dependence and has lighter tails than the Gaussian, which corresponds to a more flat shape of the contours. 

Going back to the estimated density in \autoref{fig:plot_types}, we see a rather flat shape in the lower left corner and a more spiky shape in the upper right corner. This would indicate that there is no lower, but upper tail dependence. Hence, the Gumbel copula is the most appropriate fit choosing from the parametric families in \autoref{fig:contours}. However, we also observe some asymmetry with respect to the main diagonal. This is not reflected by any of the parametric models under consideration.


\subsection{Dependence measures}

It is often useful to summarize the dependence in a single number, a \emph{dependence measure}. Many of the popular dependence measures are functionals of the copula. In fact, this property is required by R\'enyi's axioms, see \citet{schweizer1981}. Such measures can be calculated for copula density estimates fitted with \code{kdecop()}. For example, Kendall's $\tau$ can be expressed as 
\begin{align} \label{eq:spearman}
\tau = 4 \int_{[0,1]^2} C(u, v) \, c(u, v) du dv - 1.
\end{align}
There are several ways to calculate this measure for a copula density estimate. A straightforward way is to solve the integral numerically with the \pkg{cubature} package \citep{cubature}:
\begin{CodeChunk}
\begin{CodeInput}
R> library("cubature")
R> f <- function(u) pkdecop(u, fit) * dkdecop(u, fit)
R> int <- adaptIntegrate(f, lowerLimit = c(0, 0), upperLimit = c(1, 1))
R> 4 * int$integral - 1
\end{CodeInput}
\begin{CodeOutput}
[1] 0.4666633
\end{CodeOutput}
\end{CodeChunk}
This method is quite slow because two numerical integrals (one is in \code{pkdecop}) are nested.  A second way to compute $\tau$ utilizes a stochastic interpretation of \eqref{eq:spearman}. Note that $\tau = 4 \mathrm{E}[C(U, V)] - 1$, where $(U, V)$ is a random vector with density $c$. The expectation can then be approximated by Monte Carlo integration:
\begin{CodeChunk}
\begin{CodeInput}
R> uv_qmc <- rkdecop(10^4, fit, quasi = TRUE)
R> 4 * mean(pkdecop(uv_qmc, fit)) - 1
\end{CodeInput}
\begin{CodeOutput}
[1] 0.4685433
\end{CodeOutput}
\end{CodeChunk}
The third way is to compute the sample version of Kendall's $\tau$ on simulated data.
This method is asymptotically equivalent to the other methods, but simpler and faster:
\begin{CodeChunk}
\begin{CodeInput}
R> cor(uv_qmc[, 1], uv_qmc[, 2], method = "kendall")
\end{CodeInput}
\begin{CodeOutput}
[1] 0.4718781
\end{CodeOutput}
\end{CodeChunk}
We see that the three methods lead to different results, but are close. The two Monte Carlo methods have the drawback that their result is random. A large number of samples need to be drawn to get an accuracy of multiple digits, but only the first one or two digits are useful for interpretation of the measure.

Other copula-based measures of dependence can be computed similarly. The function  \\ \code{dep_measures()} automates these computations for Kendall's $\tau$, Spearman's $\rho$, Blomqvist's $\beta$, Gini's $\gamma$, van der Waerden's coefficient, Mutual Information, and Linfoot's correlation coefficient. For the definition of these measures, see, e.g., \citet{Nelsen06}, \citet{genest2005}, and \citet{joe89}. 
\begin{CodeChunk}
\begin{CodeInput}
R> dep_measures(fit)
\end{CodeInput}
\begin{CodeOutput}
   kendall   spearman  blomqvist       gini vd_waerden      minfo    linfoot 
 0.4708465  0.6527604  0.4980365  0.5368616  0.6561498  0.3277878  0.6934384 
\end{CodeOutput}
\end{CodeChunk}
\pkg{kdecopula} can be used similarly to detect other dependence properties, like positive quadrant dependence \citep[see,][]{gijbels2010positive, racine2015mixed}.

\section{Implementation based on spline interpolation}
\label{sec:splines}

Typically, the evaluation of a kernel density estimate requires going back to the original data. As a result, the computational effort increases with the sample size. We avoid that issue by evaluating the actual density estimate only once on a fixed number of grid points. For further evaluations we use cubic spline interpolation between the values on this grid. This way, the density can be evaluated efficiently --- independently of the sample size. It has the additional advantage that analytical expressions for integrals of the (interpolated) density estimate become available. We make use of that fact to implement a fast renormalization scheme that ensures that the the density estimate is close to a \emph{bona fide} copula density.

\subsection{Evaluating the estimate by cubic spline interpolation}

Recall that the support of a copula density is the unit cube $[0,1]^2$. Let $m \in \N$ and define a finite set of points $p_j \in [0, 1]$ such that $0 \le p_1 < \dots < p_m \le 1$. Then, the set 
\begin{align*}
\mathcal{P}_m = \{(u_j, v_k): (j, k)  \in \{1, \dots, m\}^2\} = \{(p_j, p_k): (j, k)  \in \{1, \dots, m\}^2\}
\end{align*} defines a symmetric grid on the unit cube. Cubic splines are piecewise cubic polynomials that can be used to approximate or interpolate some function between points on a grid. We will show how cubic spline interpolation can be used to approximate a copula density estimate. We explain in detail how a one-dimensional cubic spline interpolation is constructed  when one of the coordinates is fixed. The two-dimensional case is a straightforward extension and only sketched.


\subsubsection*{The one-dimensional case}

Let us first fix $v_k$ and assume that the values of an estimate  $\wh c(\cdot, v_k)\colon [0,1] \to \R_+$ are available on the grid points $u_j$, $1 \le j \le m.$ We want to interpolate the function $\wh c(\cdot, v_k)$ at another point $u^*$, where $u_j < u^* < u_{j+1}$ for some $j \in \{2, \dots, m-2\}$. We define the interpolated curve segment $\tilde c^{j,j+1}(\cdot, v_k)\colon [u_j, u_{j+1}] \to \R$ as some cubic polynomial
\begin{align*}
	\tilde c^{j,j+1}(u, v_k) = a_0^{j,j+1} + a_1^{j,j+1}(u - u_j) + a_2^{j,j+1}(u -u_j) ^2 + a_3^{j,j+1}(u - u_j)^3.
\end{align*} 
A cubic polynomial defined on a closed interval is fully determined by its function values and first derivatives at the boundary points. Define $\tilde c_1^{j,j+1}$ as the partial derivative of $\tilde c^{j,j+1}$ w.r.t.\ its first argument. After some simple algebraic manipulations, we find that the coefficients of a cubic spline approximation can be written as
\begin{align*}
	a_0^{j,j+1} &= \tilde c^{j,j+1}(u_j, v_k), \\
	a_1^{j,j+1} &= \tilde c_1^{j,j+1}(u_j, v_k),\\
	a_2^{j,j+1} &= -3 \tilde c^{j,j+1}(u_j, v_k) + 3\tilde c^{j,j+1}(u_{j+1}, v_k) - 2\tilde c_1^{j,j+1}(u_j, v_k) - \tilde c_1^{j,j+1}(u_{j+1}, v_k),\\
	a_3^{j,j+1} &= 2\tilde c^{j,j+1}(u_j, v_k) - 2\tilde c^{j,j+1}(u_{j+1}, v_k) + \tilde c_1^{j,j+1}(u_j, v_k) + \tilde c_1^{j,j+1}(u_{j+1}, v_k).
\end{align*}
Now we replace $\tilde c^{j,j+1}(u_j, v_k)$ and $\tilde c^{j,j+1}(u_{j+1}, v_k)$ by the known values $\wh c(u_j, v_k)$ and $\wh c(u_{j+1}, v_k)$. Similarly, we want to replace the derivatives $\tilde c_1^{j,j+1}(u_j, v_k)$ and $\tilde c_1^{j,j+1}(u_{j+1}, v_k)$ by  $\wh c_1(u_j, v_k)$ and $\wh c_1(u_{j+1}, v_k)$. These are unknown, but can approximated by a finite difference scheme. We set 
\begin{align*}
	\tilde c_1^{j,j+1}(u_j, v_k) &= \frac{\wh c(u_j, v_k) - \wh c(u_{j-1}, v_k)}{u_{j}- u_{j-1}} - \frac{\wh c(u_{j+1}, v_k) - \wh c(u_{j-1}, v_k)}{u_{j+1} - u_{j-1}} + \frac{\wh c(u_{j+1}, v_k)  - \wh c(u_j, v_k)}{u_{j+1}- u_{j}}, \\
	\tilde c_1^{j,j+1}(u_{j+1}, v_k) &= \frac{\wh c(u_{j+1}, v_k) - \wh c(u_{j}, v_k)}{u_{j+1}- u_{j}} - \frac{\wh c(u_{j+2}, v_k) - \wh c(u_{j}, v_k)}{u_{j+2} -u_{j}} + \frac{\wh c(u_{j+2}, v_k)  - \wh c(u_{j+1}, v_k)}{u_{j+2} - u_{j+1}}.
\end{align*}
Note that these can only be computed for $2 \le j \le m-2$, since four distinct values $\wh c(u_{j+j^*}, v_k)$, $j^* = -1, 0, 1, 2,$ show up in the above formulas. For fixed $v_k$ and some $u \in [u_2, u_{k-1})$, the spline approximation $\tilde c(u, v_k)$ of the function $\wh c(u, v_k)$ can then be written as
\begin{align*}
	\tilde c(u, v_k) = \sum_{j=2}^{m-2} \tilde c^{j,j+1}(u, v_k) \ind_{[u_j, u_{j+1})}(u).
\end{align*} 
We extended this to allow for the full range ($u \in [0,1]$) by extrapolating the `outer' two  polynomials at the borders, i.e.,
\begin{align*}
	\tilde c(u, v_k) = \sum_{j=2}^{m-2} \tilde  c^{j,j+1}(u, v_k) \ind_{A_j}(u), \qquad \mbox{where}\quad 
	A_j = \begin{cases}
	[0, u_{3}) & \mbox{for } j = 2, \\
	[u_j, u_{j+1}) & \mbox{for } 3 \le j \le k-3, \\
	[u_{m-2}, 1] & \mbox{for } j = m-2.
	\end{cases}
\end{align*} 
The advantage of cubic spline interpolation is that it is easy to compute. In particular, the computational effort only depends on $m$, the number of knots. Additionally, the above approximation allows to write integrals as a sum of quartic polynomials, which can be computed  equally fast. This will prove advantageous in \autoref{subsec:renorm}, where we use such integrals to renormalize the copula density estimates.

\subsubsection*{The two-dimensional case}

\begin{figure}
\centering
	\subfloat[]{\includegraphics[width=0.4\textwidth]{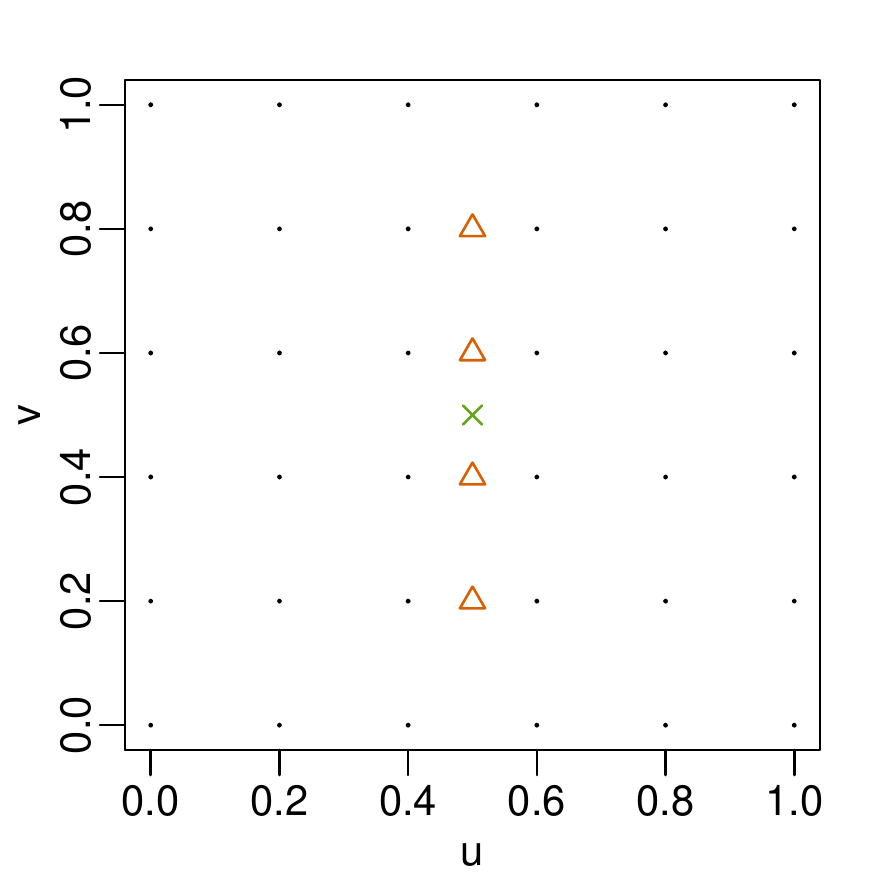}
    \label{interp_fig}
    } 
    \hspace{0.8cm}
	\subfloat[]{\includegraphics[width=0.4\textwidth]{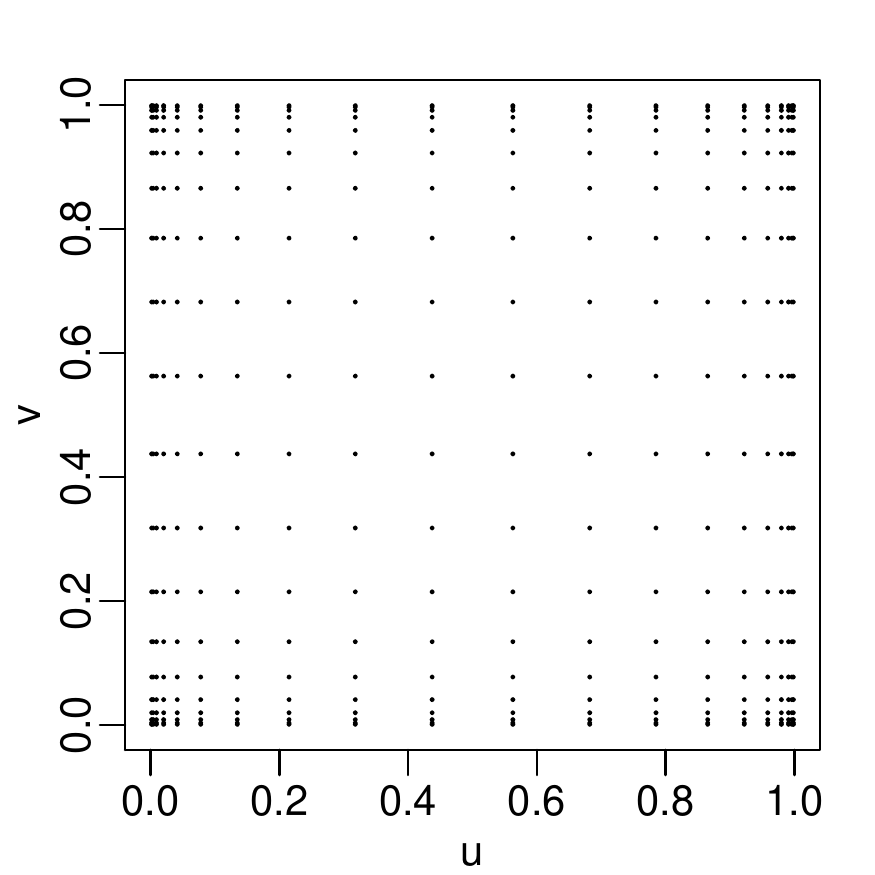}
    \label{grid_fig}
    }
    \caption{(a) Visualization of a two-dimensional interpolation as the sequence of two one-dimensional interpolations. (b) A grid with 20 knots that is equally spaced after transformation with the inverse Gaussian $cdf$.}
\end{figure}

Bivariate functions can be approximated similarly by a sequence of two one-dimensional interpolations. We will illustrate this by a small example and omit any further details \citep[for more, cf.,][]{Habermann07}. \autoref{interp_fig} shows the unit cube with grid points $(u_j, v_k) = (j/5, k/5)$, $j,k = 0, \dots, 5$, indicated as dots. Assume we know all function values $\wh c(u_j, v_k)$ on this grid and want to approximate the function at the point $(0.5, 0.5)$ indicated by a cross. We first do four one-dimensional (horizontal) interpolations $\tilde c(0.5, v_k)$, $j = 1, 2, 3, 4$ (triangles). Note that all values that are required to calculate the spline coefficients are known. Another one-dimensional (vertical) interpolation based on the four new values $\tilde c(0.5, v_k)$ gives us the final interpolated value $\tilde c(0,5, 0.5)$. 

\subsubsection*{Choice of grid}

In \autoref{interp_fig} we showed a grid that has equal spacings between grid points. This seems natural, but in our context we found it more appropriate to use a grid that is equally spaced after a transformation by the inverse Gaussian $cdf$. \autoref{grid_fig} depicts such a grid with 20 knots. It was constructed by placing $20$ equidistant knots on the line segment $[-3, 3]$ and then applying the Gaussian $cdf$ to them. The two-dimensional set-product of these 20 points yields the final two-dimensional grid. 

We see that the grid points are more sparse in the center of the unit square and concentrate towards the boundaries and corners. This choice takes into account that for copula densities the areas near the corners are most important. In those areas, copula densities often explode while being rather flat in the center. This allows us to keep the approximation errors in the important areas small. A nice side-effect is that the marginal normal contour plots described in the last section can be visualized more nicely.  The number of knots can be controlled by the \code{knots} argument of \code{kdecop} and defaults to 30. A smaller number reduces computation time, but comes at the cost of a larger approximation error.

\subsection{Renormalization of the density estimate} \label{subsec:renorm}

We now introduce the idea of iterative renormalization of kernel copula density estimators. Recall that by the definition of a copula, the marginal densities have to be uniform, i.e.,
\begin{align}
\int_0^1 c(u, s) ds = \int_0^1 c(s, u) ds = 1, \qquad \mbox{for all } u \in [0,1]. \label{renorm:unifmargs_eq}
\end{align} 
This property is of particular importance, when other functionals of the density are of interest. For example, assume that we integrate the density estimate to obtain an estimate for the corresponding conditional \emph{cdf}, $C(v|u) = \int_0^v c(u, s) ds$. This is a common task in vine copula models which gained a lot of popularity following the seminal paper of \citet{Aas09}. If the estimated density does not satisfy the uniform margins property, the estimate of the conditional \emph{cdf} may exceed unity, which makes it problematic. The lack of uniform margins was mostly ignored in the literature, although kernel estimates usually do not satisfy the \emph{uniform margins property} \eqref{renorm:unifmargs_eq}.

Now let $\wh c(u,v)$ be a consistent kernel estimator of $c(u,v)$ for all $(u,v) \in [0,1]^2$.  From Sklar's theorem for density functions \eqref{introduction:sklar_eq}, 
we know that dividing a bivariate density by the product of its marginal densities yields a copula density. Hence, a simple way to adjust the estimator is to divide the initial estimate by $\int_0^1 \wh c(u, s) ds \int_0^1 \wh c(s, v) ds$. The renormalized estimator writes
\begin{align}
\wh c^*(u, v) = \frac{\wh c(u,v)}{\int_0^1 \wh c(u, s) \int_0^1 \wh c(s, v) ds}.
\label{renorm:renorm_eq}
\end{align}
and is asymptotically equivalent to the original one under mild conditions. For sophisticated kernel estimators --- such as the beta kernel or local likelihood estimators --- the two integrals have to be computed numerically. Conveniently, the spline approximations introduced in the previous section allow for fast computation of the integrals in \eqref{renorm:unifmargs_eq}. 

The proposed renormalization procedure can be split into two steps. 
\begin{enumerate}[1.]
	\item Find a spline approximation of the initial estimate that is defined by its values on a finite grid.
	\item Renormalize the approximated density values on this grid by dividing by the (approximated) marginal densities (see \autoref{renorm:renorm_eq}).
\end{enumerate} 

The resulting estimate will typically be closer to a \emph{bona fide} copula density. However, the renormalization is only carried out on a finite number of grid points. Apart from that grid, the renormalized estimate typically still does not satisfy the uniform margins property. But we can simply repeat the two steps above until a satisfactory result is achieved. Our experience suggests that a very small number of iterations is sufficient. The number of iterations can be set by the \code{renorm.iter} argument of \code{kdecop} and defaults to three.

The renormalization will turn out to have two benefits: functionals of kernel estimates will show the desired behavior and, additionally, the estimates are more accurate (see \autoref{sec:simulations}). For the latter there is an intuitive interpretation. By ensuring that margins are uniform, we incorporate additional information about the true density. This reduces the set of plausible estimates and increases the probability of being `close' to the true one.

\section{Comparison of methods}
\label{sec:simulations}

We compare all estimators implemented in this package in a simulation study. Additionally, we include results for other nonparameteric copula density estimators available in \proglang{R}.


\subsection{Setup}

We consider the following estimators:
\begin{itemize}
	\item MR, beta, T, TLL1, TLL2, TLL1nn, TLL2nn TTPI, TTCV: These estimators are provided by the \code{kdecop} function presented in this paper (cf., \autoref{sec:review}). 
    \item np: The kernel estimator provided by the \code{npcopula} function in the \pkg{np} package \citep{np}.
    \item ks: The kernel estimator provided by the \code{kcopula.de} function in the \pkg{ks} package \citep{ks}.
	\item bern, bspl: The penalized Bernstein polynomial and B-spline estimators provided by the \code{paircopula} function in the \pkg{penDvine} package \citep{penDvine}.
    
\end{itemize}
Default settings are used for all implementations. In particular, smoothing parameters are selected automatically.

As a performance measure, we use the \emph{integrated absolute error (IAE)}
\begin{align*}
	\mbox{IAE}\bigl[\wh c(u, v)\bigr] = \int_{[0,1]} \int_{[0,1]} \bigl\vert \wh c(u, v) - c(u, v) \bigr\vert du dv,
\end{align*}
where we estimate the integrals as the mean over the grid $(j/101, k/101)$, $j, k = 1, \dots, 100$.

\begin{figure}
\centering
\subfloat[Mixture I]{
\includegraphics[width=0.33\textwidth]{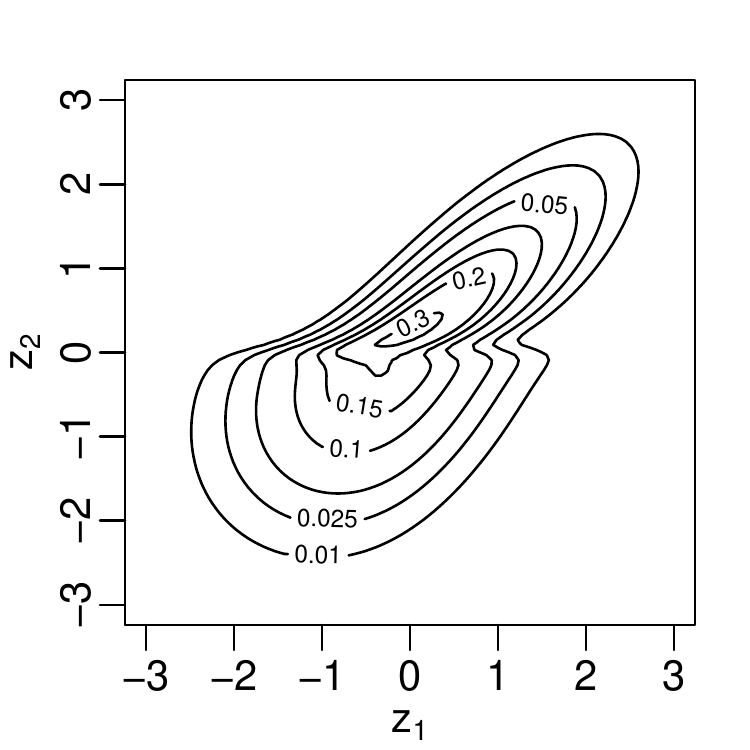}
}
\subfloat[Mixture II]{
\includegraphics[width=0.33\textwidth]{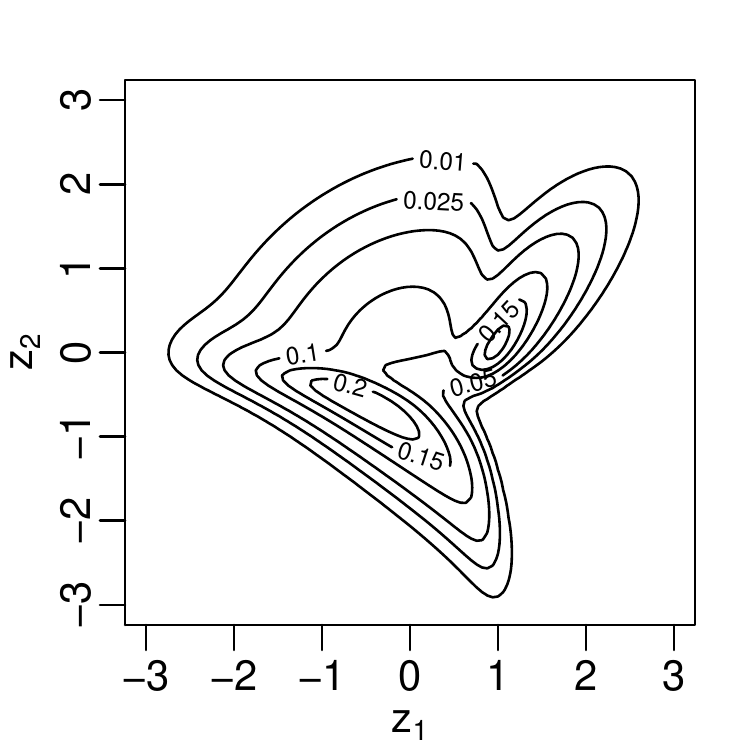}
}
\subfloat[Mixture III]{
\includegraphics[width=0.33\textwidth]{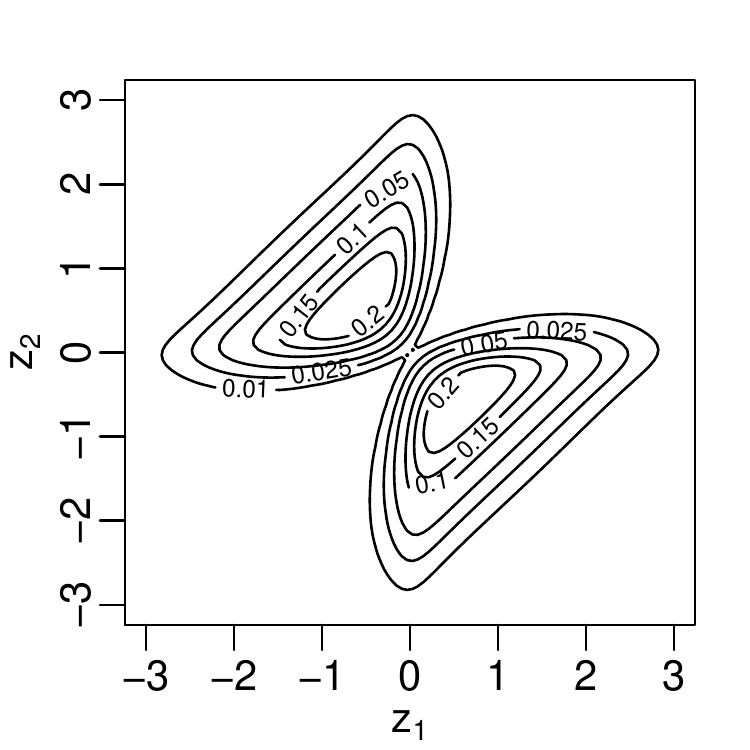}
}
\caption{Marginal normal contour plots of copulas derived from mixtures of multivariate Gaussian distributions.}
\label{fig:mixtures}
\end{figure}

\begin{figure}
\centering
\includegraphics[width = 0.9\textwidth]{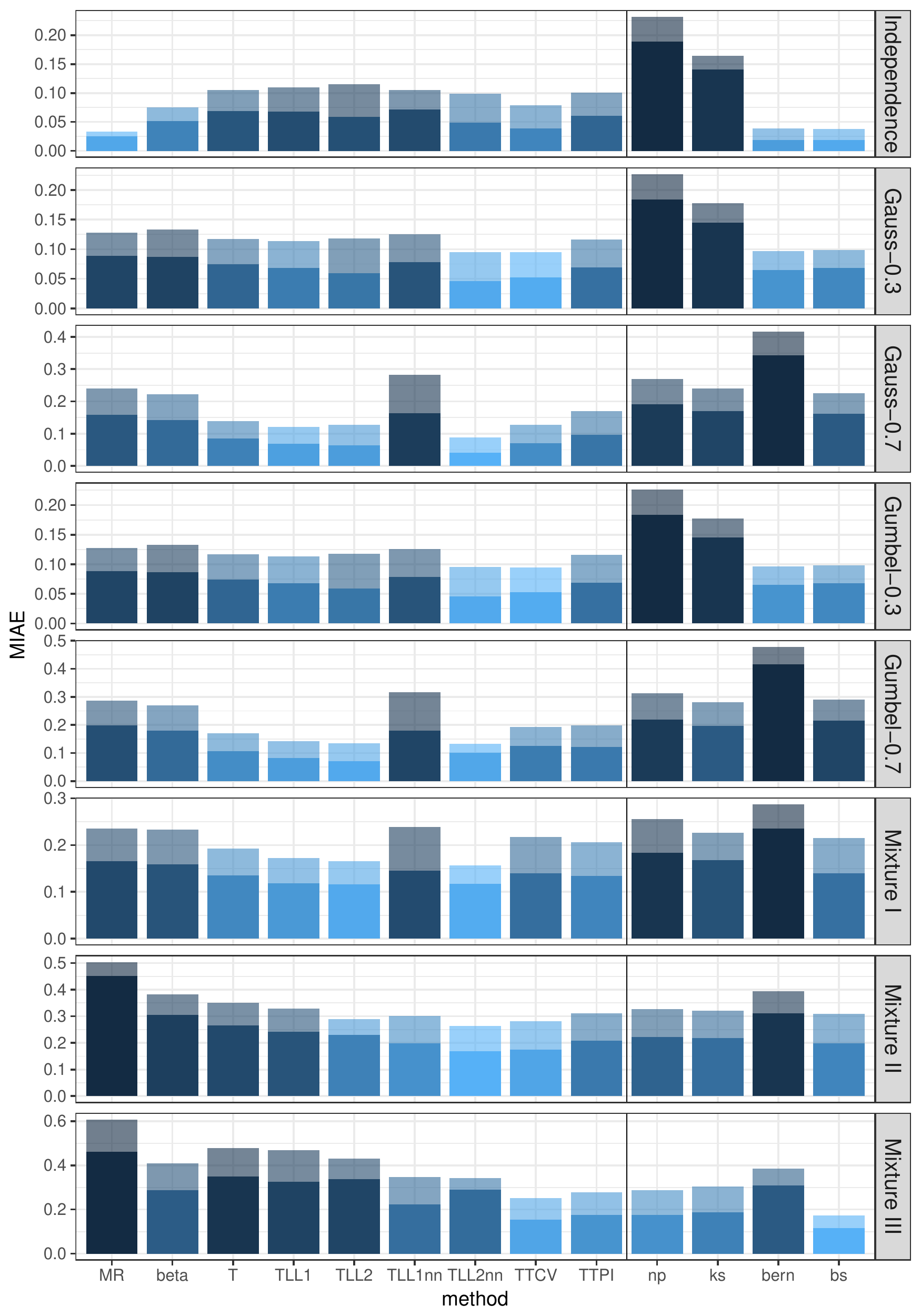}
\caption{Mean of the integrated absolute error achieved by all methods in different scenarios. Each bar is split in two parts, where the larger corresponds to $n = 200$ and the smaller  to $n=1\,000$.}
\label{fig:sim}
\end{figure}

We study several target copula densities and two sample sizes ($n = 200, n = 1\, 000$). We use three families from \autoref{fig:contours} (Independence, Gaussian, Gumbel) and three copulas derived from multi-modal mixtures of bivariate Gaussian distributions (marginal normal contour plots are shown in \autoref{fig:mixtures}). For the Gaussian and Gumbel copulas we create  scenarios with weak and strong dependence (Kendall's $\tau$ of 0.3 and 0.7). For mixture families we use the package \pkg{GMCM} \citep{bilgrau2016gmcm}.

\subsection{Error analysis}

\autoref{fig:sim} shows the mean of the IAE over $100$ replication for various scenarios. The figure is divided into 8 panels, each representing a different target density. Each panel shows results for the nine estimators implemented in \pkg{kdecopula} (left of the dotted line) and four estimators implemented in other packages on CRAN. Each estimator is represented by a bar which is split in two parts. The solid part indicates the mean IAE for $n=1\,000$, the transparent part indicates the mean IAE for $n=200$. As expected, all estimators improve when the sample size increases.

Two scenarios, the independence and Mixture III copulas, show somewhat different results from the others. We shall discuss them in more detail later. In all other scenarios, TLL2nn is the top performer and TLL2 is a close second. In general, the transformation methods seem to work better than MR and beta. The estimators from the \pkg{np} and \pkg{ks} packages are often among the least accurate. The methods from the \pkg{penDvine} package work rather well when there is weak dependence, but struggle when dependence is strong. The relative performance is largely consistent across the two sample sizes.

The density of the independence copula is constant. The two estimators from \pkg{penDvine} perform best for this target. The reason is that their basis function formulation can reproduce constants, so they behave similar to a parametric estimator in this case. The independence copula is an example where the transformation approach is suboptimal. A constant density can be estimated easily because there is no curvature. But after transformation, the target density is a bivariate Gaussian which is more difficult to estimate. 

The second scenario standing out is where Mixture III is the target density. The B-spline estimator from \pkg{penDvine} is most accurate, followed by TTCV and TTPI. This scenario is the only one where the other transformation methods are consistently outperformed by non-\pkg{kdecopula} methods. The issue here is how the bandwidth matrix is selected. Methods T, TLL1, TLL2, TLL1nn, and TLL2nn all use a bandwidth matrix that is proportional to square root of the empirical covariance. This is often a good choice because it `stretches' the kernels in a way that resembles the shape of the data. For Mixture III, the overall correlation is negative. The correlation in each mixture component, however, is strongly positive. In this case the shape of the kernels does not reflect to the local orientation of the data. TTCV, TTPI, and non-\pkg{kdecopula} methods do better in this scenario because they do not rely on such a rule of thumb, but use more sophisticated criteria. This comes at a cost in terms of speed, as we shall see in \autoref{sec:time}.


\subsection{The effect of renormalization}

In \autoref{subsec:renorm} we claimed that the renormalization algorithm implemented in this package improves the performance of the estimators. The results presented in \autoref{fig:sim} are based on default settings, i.e., three iterations of the algorithm. \autoref{tab:sim2} show the relative reduction of IAE compared to non-normalized estimators averaged over all scenarios.

\begin{table}[th]
\centering
\begin{tabular}{rrrrrrrrrr}
 MR & beta & T & TLL1 & TLL2 & TLL1nn & TLL2nn & TTCV & TTPI \\ 
  \hline
 13\% & 15\% & 25\% & 39\% & 21\% & 28\% & 29\% & 24\% & 22\% \\ 
\end{tabular}
\caption{The effect of renormalization: numbers indicate by how much the IAE could be improved after three iterations of the renormalization algorithm (rounded to the next integer).}
\label{tab:sim2}
\end{table}

We observe that the performance has improved for all estimators. The average gain ranges between 13\% and 39\%. This contributed significantly to the good performance observed in \autoref{fig:sim}. In fact, the estimator that could be improved the second most is TLL2nn, the top performer in our initial study.


\subsection{Computation time} \label{sec:time}

Some design choices in \pkg{kdecopula} prioritize speed over accuracy. One example is the use of spline interpolation, another is the choice of bandwidth selection methods. \autoref{tab:time} shows the computation time for most methods. Methods TLL1 and TLL1nn are omitted to make the table more compact; they are slightly faster than their `nn'-variants. 

\begin{table}[ht]
\centering
\begin{tabular}{r|rrrrrrrrrrr}
 n & MR & beta & T & TLL2 & TLL2nn & TTCV & TTPI & np & ks & bern & bs \\ 
  \hline
 200 & 0.11 & 0.26 & 0.07 & 0.31 & 0.62 & 3.76 & 0.35 & 3.03 & 4.61 & 1.88 & 2.10 \\ 
  1000 & 0.37 & 0.99 & 0.11 & 0.68 & 1.30 & 58.10 & 13.63 & 37.84 & 41.96 & 3.08 & 3.45 \\ 
\end{tabular}
\caption{Computation time (in seconds) required for estimating the density and evaluating it on a $100 \times 100$ grid.}
\label{tab:time}
\end{table}

All \pkg{kdecopula} methods except TTCV and TTPI are much faster than estimators from other packages. The difference gets larger for larger sample sizes. Note that TTCV, TTPI, np, and ks are much slower when $n=1000$. This is caused by them using sophisticated bandwidth selection techniques that have large complexity with respect to sample size.

\section{Summary and extensions}
\label{sec:conclusion}

We have described the \proglang{R} package \pkg{kdecopula}, which implements several cutting-edge kernel estimation techniques for copula densities. The package allows for automatic selection of the smoothing parameter and resampling. Several plotting options make it particularly useful for the exploratory analysis of copula data. Its abilities have been illustrated by small code examples. 

The implementation utilizes spline interpolation for fast evaluation and renormalization of the density estimates. Simulations show that the implementations in this package perform best among available methods for nonparametric copula density estimation. A contributing factor for the good performance is the renormalization of the estimators, which was shown to notably improve the accuracy.

Compared with the functionality provided by the \pkg{np} package, \pkg{kdecopula} lacks two features: 
\begin{enumerate}
\item It does not provide functionality for estimating copula densities when the data contain discrete variables. A problem is that the copula is not unique when margins are discrete, but there are infinitely many copulas that, when combined with the marginal distributions, lead to the same joint distribution \citep[see, e.g.,][]{genest2007primer}. It is unclear which of these copulas shall be estimated and how to justify the choice. Additionally, estimating a copula from discrete data necessarily involves modeling of the marginal distributions, which is deliberately avoided in \pkg{kdecopula}.
\item It does not allow for more than two variables. One major issue is that \pkg{kdecopula}  uses interpolation to evaluate and renormalize the estimators. In more than two dimensions the number of grid points explodes rapidly and renders the interpolation approach infeasible. 
\end{enumerate}

A \pkg{kdecopula}-based solution for both points is the \pkg{kdevine} package \citep{kdevine}. It implements a kernel estimator of general multivariate densities based on vine copulas \citep{nagler2016evading}, which use marginal densities and bivariate copulas as building blocks. \emph{Continuous convolution} \citep{nagler2017generic} is used to handle discrete variables, which induces copulas similar to the multilinear copula \citep[see,][]{genest2007primer,genest2014empirical}.

\section*{Supplementary material}

R code for the simulation study can be found at:
\begin{center}
\href{https://gist.github.com/tnagler/bd194a711026c3d375ab6ae023a5bad5}{https://gist.github.com/tnagler/bd194a711026c3d375ab6ae023a5bad5}.
\end{center}

\section*{Acknowledgments}

This work was partially supported by the German Research Foundation (DFG grant CZ 86/5-1). The author wants to thank two anonymous referees for many remarks and suggestions that considerably improved the article and software.

\bibliography{article}

\end{document}